**Functional Imaging of Conceptual Representations**

*Introduction*

In philosophy, a concept is defined as "an idea or mental picture of a group or class of objects formed by combining all their aspects" (Oxford American Dictionary). Because access to a concept can be gained through multiple modalities such as vision, audition and somatosensation, one critical question is how the brain supports a unified representation of a concept. One hypothesis states that such representations depend upon multimodal binding processes within a single region. It has been proposed that such a region is contained within the temporal lobe (Patterson 2007). Imaging work has implicated various candidate regions include temporal gyrus (Rissman et al. 2003) fusiform gyrus (Wheatley et al. 2005) and perirhinal cortex (Taylor et al. 2006). To characterize the role of these regions, we examined behavioral priming and physiological suppression in response to repetitions of concepts either within or across perceptual modalities.

When observed in response to exact perceptual repetitions, behavioral priming is believed to reflect facilitated processing for the repeated presentations. Similarly, blood-oxygenation-level-dependent (BOLD) suppression within a region is believed to reflect facilitated processing for repetitions of the features or operations represented within that region. Consequently, differential BOLD suppression effects within perceptual regions such as visual cortex have been used to identify regions that preferentially process particular subcategories of visual stimuli such as objects (e.g. Epstein and Kanwisher 1998, Koutstaal et al. 2001, Ewbank et al. 2005). However in the case of perceptually *distinct* (cross-modality) item repetitions (e.g. an item presented first as a picture and then repeated as spoken word), any facilitated processing must be driven by factors other than the strict perceptual repetition of stimulus features. In other words, cross-modality priming reflects repetition of a conceptual item representation. By the same token, if a region exhibits suppression for cross-modality repetitions, one could infer that the region is involved in processing aspects of the stimulus that are not strictly tied to the current percept, aspects which rather are more conceptual in nature. Moreover, if across many different conditions, the magnitude of suppression correlated with the magnitude of behavioral priming, one might further deduce that the representation in this region denotes at least some component of the conceptual information underlying behavioral priming.

Suppression effects in certain regions may reflect the causative neural mechanisms that contribute to behavioral priming effects (Henson 2003). Indeed, disrupting suppression, either chemically or through transcranial magnetic stimulation, has been shown to correlate with disruptions in observed behavioral priming (Thiel et al. 2001, Thiel et al. 2005, Wig et al. 2005). However, not all suppression effects are equal. Effects can vary across regions and in some brain regions suppression occurs independently of behavioral priming (e.g. Sayres and Grill-Spector 2006). Typically, examinations of the relationship between behavioral priming and suppression query the extent to which suppression effects in different brain regions correlate with single measures of behavioral priming. In contrast, we were interested in the extent to which suppression effects within a *single* brain region would correlate with *multiple* measures of behavioral priming. This measure targets regions that contribute to item processing encompassing multiple percepts, and identifies regions whose suppression effects are most predictive of behavioral priming. Hence, this method can identify regions with item representations most similar to the unified conceptual representation. In the present study, items were first presented in one of three distinct modalities: spoken words, written words and pictures (Figure 1a). On each presented trial, response time was measured as participants make a semantic judgment (natural or manmade) for the item. After one to four intervening trials, each item was repeated, in either the same or a different modality. Perceptually distinct (cross-modality) item repetitions were used to examine activation of a single concept through multiple perceptual routes. Item repetition effects were indexed with behavioral priming. To address our hypothesis BOLD signal was queried for repetition suppression effects (suppression) that correlated with behavioral priming effects. Perirhinal cortex, a region in medial temporal lobe, emerged as the region with item representations most closely linked with conceptual representations.

## Results

*Behavioral.* We measured response times for each trial, and analyzed all first presentations of items separately from second presentations (Figure 1). There were three groups of first presentation trials: spoken words, written words, and pictures. Since presentation times varied across presentation modalities (notably, for both first and second presentations, response times to auditory presentations took longer than visual ones) the three groups were analyzed separately. As the repetitions were presented in a fully crossed design, the three modalities ultimately gave rise to nine repetition conditions. The second presentations of each modality (3) were further subdivided by the modality of the previous item presentation, creating three groups of trials within each of the three modalities: totaling the nine repetition conditions according to first and second presentation modality. Mean response times (RTs) were calculated separately for each of the three groups of first presentations and the nine groups of second presentations. These calculations were performed for each of the 16 subjects (Figure 1). Across all participants, for first presentations, the mean RT for spoken words was 1243.7 ms, for pictures, 777.8 ms, and for written words 787.6 ms (Figure 1b). To assess priming, the mean RT of each second presentation group was compared to the mean RT of first presentations in the corresponding modality (e.g. for the grouped second presentation of items that were first presented as pictures and later repeated as spoken words, the mean RT (1192 ms) is compared to that of spoken word first presentations (1244 ms) for a reduction of 52 ms or about 4.2%).

Comparison of first and second presentation RT means reveals significant priming effects (with differing magnitudes) for each of the nine conditions (one tailed t-tests, all $p$s < .008)(Figure 1b). The magnitude of priming for the within-modality repetitions was significantly greater than that of the cross-modality repetitions (one tailed t-tests, all $p$s < .005). Conditions with picture presentations had the least robust cross-modality priming effects. Within cross-modality repetitions, priming was greater for the spoken word second presentation when the first presentation had been a written word than when it had been a picture ($p < .05$). Similarly, for written word second presentations, priming was greater when the first presentation had been a spoken word, than when it had been a picture ($p < .05$).

*Imaging.* We created statistical parametric maps to identify regions that showed greater activation for the first presentation of items compared to the second presentation (suppression). Separate linear contrast for each of the three within-modality repetition conditions (spoken-spoken, picture-picture, written-written) revealed within-modality suppression effects in various frontal and temporal lobe regions, consistent with extant literature on perceptual repetition effects (Supplementary Figure 1). To target conceptual processing, we queried the brain for regions showing both within-modality AND cross-modality repetition suppression. Specifically, we used a conjunction analysis (Friston et al. 1999) to reveal brain regions showing reduction in activity compared of the second presentation compared to the first presentation for four repetition conditions: 1) spoken words - spoken words, 2) written words - written words 3) spoken words – written words 4) written words – spoken words (joint $p = .005$). Because the cross-modality conditions involving pictures had less robust behavioral priming, and thus we would predict differential power within the associated suppression effects, they were not included in this conjunction. Cortical areas revealed by this analysis include regions previously implicated in conceptual processing (inferior frontal gyrus (-54 27 3), perirhinal cortex (-36 -21 -27), fusiform gyrus (-36 -33 -27), superior temporal sulcus (60 0 9)). We created functional regions of interest (ROIs) from these clusters and, using deconvolution analyses, examined the pattern of suppression across all nine conditions. Within each ROI, we calculated the difference in mean peak BOLD response for first and second presentations, in a manner parallel to the behavioral analyses.

Suppression data provide further support for the proposed involvement of these regions in conceptual processing. In inferior frontal gyrus, we observed significant suppression for all nine conditions ($p < .05$). In perirhinal cortex, significant suppression ($p < .05$) was seen for eight of the nine conditions, with the ninth condition, spoken words – pictures showing a marginally significant effect ($p = .07$) (Figure 2b). However, in fusiform cortex, suppression effects were significant in only the five conditions with the most robust behavioral effects: 1) spoken words - spoken words 2) written words - spoken words 3) pictures – pictures 4) spoken words – written words, written words – written words. (Figure 2d). Further suppression in superior temporal sulcus is significant only for three conditions: spoken words - spoken words, written words - pictures, and spoken words - pictures.

To assess the extent to which the suppression within a region reflected the behavioral priming, we correlated the group behavioral priming data with the group BOLD suppression across all nine (9) conditions. Significant correlations were observed in only two temporal lobe clusters: perirhinal cortex and fusiform gyrus (r = .83 $p < 0.0004$.) (Figure 3a,b). In contrast, while regions such as superior temporal sulcus (60 0 9) and inferior frontal gyrus (-54 27 3) also showed conceptual suppression effects, suppression in these regions did not significantly correlate with behavioral priming. ($ps > 0.4$, Supplementary Figure 2,3).

*Discussion*

We observed significant suppression for both within and cross-modal repetitions in perirhinal cortex, fusiform gyrus, inferior frontal gyrus, and superior temporal sulcus demonstrating that these areas support multi-modal item representations and, thus may be critical for conceptual processing.

From the behavioral data we see that while conceptual repetitions lead to behavioral priming, the stimulus percept influences the magnitude of this priming (see also Holcomb et al. 2005). Even within second presentations of the same modality, the size of the priming effect varies according to the modality of the first presentation. These effects suggest that the active representation is a function of both the immediate and prior perceptual cues. Also, the increased priming effects for perceptually matching repetitions, provide evidence favoring theories which propose that priming consists of two components, one purely perceptual, arising from automatic processing of the external stimulus, and an additional conceptual one, reflecting internal mnemonic contributions (Henson 2003).

From the BOLD data, we can discern that in contrast to effects in superior temporal sulcus and inferior frontal gyrus, suppression effects in regions such as perirhinal cortex and fusiform gyrus significantly correlate with subjects' behavioral pattern of priming. Effectively, in these areas suppression is modulated according to stimulus modality in a similar manner as the behavioral priming. Interestingly, unimodal studies of conceptual processing have found that suppression in perirhinal cortex and fusiform gyrus can be eliminated by changing the behavioral task (Dobbins 2004; O'Kane et al. 2005). These findings reveal that active representations in these areas are modulated by task, and providing additional evidence that representations in these areas encompass features other than the stimulus cue.

Extant literature on perceptual repetition in humans also reveals suppression in some regions that is dissociated from behavioral priming; temporary lesion studies with perceptually identical repetitions of visual stimuli find that although suppression is observed in many regions including visual cortex, behavioral priming is more closely dependent on suppression in the frontal lobe than visual cortex (Thiel et al. 2005; Wig et al. 2005). In fact, utilizing transcranial magnetic stimulation to eliminate suppression is more disruptive to behavioral priming when frontal, not visual, regions are targeted (Thiel et al. 2005; Wig et al. 2005). While one cannot conclude that suppression in frontal cortex causes behavioral priming, the data do suggest that items are represented in multiple regions, and that even in a task of visual perception, the representation required for behavioral priming is not necessarily the one in visual cortex. In the present data we observe significant conceptual suppression in frontal lobe regions, suggesting that they are involved in conceptual processing. However, when we compare BOLD data and behavioral data utilizing all the conditions available within the design, suppression in many of these regions (such as inferior frontal gyrus) does not directly correlate with the observed behavioral priming effects. Since the item representations in these regions differ from the representation reflected by the priming data, what do processes in these regions contribute to conceptual representation? To answer this question, we must first consider how processes in multiple regions contribute to perceptual representations.

Recent vision studies have found that perceptual representations are not only a function of purely stimulus-driven, feed-forward processes, but also depend on recurrent feedback between sensory cortices and other regions (Bressler et al, 2008, Supèr et al. 2001, Grelotti et al. 2005, Vuilleumier et al. 2004). Such recursive inputs can be observed in sensory cortices as activations that reflect percepts rather than external stimuli. As such, it follows that conceptual representations may also require recurrent processing across regions. Work in non-human primates has demonstrated that perirhinal cortex receives inputs from a variety of cortical regions including visual temporal regions, superior temporal sulcus, and frontal lobe (Suzuki and Amaral 1994). These connections are mostly reciprocal (although perirhinal cortex outputs additional projections to some temporal regions from which it does not receive inputs) (Lavenex et al. 2002). In the present data, the observed pattern of activity within perirhinal cortex may denote summation of iterative activity between perirhinal cortex and other regions. If so, multiple activated representations in disparate regions may give rise to a unified, active conceptual representation in perirhinal cortex.

This theory is corroborated by human electrophysiological studies of the N400 event-related-brain potential. The N400, a negativity that occurs near 400 ms, can be attenuated by repetition and has been shown through intracranial recordings in humans to originate from anterior MTL (Smith et al. 1986). Furthermore, examination of cross-modality repetitions finds that as in the present fMRI data, the perceptual modalities of repeated presentation differentially modulate the magnitude of suppression (Holcomb et al. 2005). In the linguistic literature, suppression of the N400 is theorized to reflect semantic integration (Deacon et al. 2000, Holcomb 1993, Rugg 1990), while in the memory literature, attenuation of this potential has been interpreted as a familiarity signal (Henson et al. 2003, Fernandez and Tendolkar 2006). Both interpretations are consistent with the hypothesis that activity in this region is engaged in item representation beyond simple perceptual processing.

In sum, these data support the hypothesis that perirhinal cortex is critically involved in conceptual processing. Given the connectivity of this region, activity in perirhinal cortex may reflect integration of multiple item representations across diverse regions. Accordingly, this integrated representation assessed in perirhinal cortex significantly correlates with the representation assessed by behavioral priming. In fusiform gyrus, despite the lack of significant suppression in some conditions, the *pattern* of activation similarly correlates with behavior. This suggests that fusiform gyrus may be one region that receives integrated outputs from perirhinal cortex. However, reciprocal connections between perirhinal cortex and other brain areas may enable observations of portions of the integration process in regions providing recursive inputs.

*Methods*

*Participants.* Twenty (11 female) right-handed, native English speakers participated. Informed consent was obtained in writing under a protocol approved by the Institutional Review Board of New York University. All participants reported normal vision and hearing, and no history of neurological or psychiatric disorders. We removed three participants from inclusion in all analyses due to technical difficulties with stimulus presentation. An additional participant was removed for poor behavioral performance (below criterion 80% correct trials).

*Stimuli*. We presented 285 items (234 objects, 51 scenes) twice each. While scenes were always presented as full color photographs (IMSI MasterClips© and MasterPhotos™ Premium Image Collection, 1895 Fransisco Blvd., East, San Rafael, CA 94901-5506, USA), objects could be presented in any of three modalities: spoken words, written words, and pictures. Pictures were full color photographs collected from two online photography databases (www.photoobject.net, www.cepolina.com). In a separate phase, a different cohort of participants was presented the photographs and instructed to label each with the first word that came to mind. Only photographs that received the same label across 80% of participants were included in the present study. Furthermore, these labels were in turn the basis of the spoken and written word stimuli. Spoken words were presented using wav files collected from the LDC American English Spoken Lexicon (http://www.ldc.upenn.edu/cgi-bin/aesl/aesl). Stimulus presentations for visual stimuli were fixed at 250 ms; auditory presentations varied between 500 and 1500 ms. Object stimuli were grouped into 9 sets of 26 items each. Each set was presented in one of the 9 repetition conditions, and the exact mapping was counterbalanced across subjects.

*Task.* Participants were instructed to indicate, by making a button press as quickly as possible, whether the presented item was natural or manmade. Trials were presented serially with repetitions separated by 1- 4 (mean 2.5) intervening trials. During jittered inter-trial intervals (Dale 1999), participants were instructed to fixate on a blue fixation cross which flashed green at the onset of each trial.

*Imaging Parameters.* Imaging data were collected with a 3 Tesla Siemens Allegra scanner. We acquired functional data across three scans each containing 297 volumes (TR = 2000 ms, TE = 30 ms, flip angle = 85, 35 slices, 3 x 3 x 3 mm voxels, 20% distance factor) using coronal slices angled perpendicular to the long axis of the hippocampus. The first four volumes, collected for stabilization purposes, were discarded. A high-resolution, T1-weighted, full brain, anatomical scan (magnetization-prepared rapid-acquisition gradient echo) was collected for visualization.

## Data analysis

*Behavioral Procedures.* Priming was assessed within individual participants. Only pairs of trials for which participants made correct response to both presentations of the item were included. Furthermore, trials with outlying response time (RTs more than twice the interquartile range from a median RT calculated separately for the first and second presentations of each modality) and their pairs were removed.

For first presentations, mean RTs were calculated for each of the three modalities. For second presentations, mean RTs were calculated separately for each of the nine repetition conditions. Difference scores for the nine conditions were calculated by subtracting the mean RT for the second presentation from the mean RT of the first presentation of the same modality. For comparisons across modalities, percent reduction was calculated by dividing the difference score by the mean of the first presentation of the matching modality. Across participants, one tailed paired t-tests of differences between first and second presentations were used to determine significance of priming effects for each condition.

*Imaging.* For preprocessing and analysis of fMRI data, we used SPM2 (http://www.fil.ion.ucl.ac.uk/spm), a MATLAB-based analysis package (http://www.mathworks.com).

Volumes were corrected for different times of slice acquisition, and were realigned correcting for subject movement and scanner drift. Data were next normalized to a Montreal Neurological Institute reference brain (MNI; Montreal Neurological Institute, Montreal, Canada), and then smoothed with an isotropic 6 mm full width half-maximum Gaussian kernel.

Using the general linear model implemented in SPM2, statistical parametric maps (SPMs) were first computed for individuals. The fixed-effect participant-specific estimates were entered into a second-level random-effects analysis (one-sample T-test). Regions consisting of at least five contiguous voxels that exceeded threshold were considered reliable. Functional regions of interest (ROIs) were defined with the clusters from the SPMs. We extracted ROI deconvolution data utilizing the MarsBaR (http://marsbar.sourceforge.net) software package. Further statistical analyses were performed with customized MATLAB scripts. For suppression comparisons, difference scores and percent reductions were calculated using the numerical peak of the hemodynamic responses of each subject. Calculations paralleled those of the condition RT means of the behavioral data. Behavioral-BOLD correlations were assessed using Kendall's Tau.

Acknowledgements

Figures
**Figure 1**

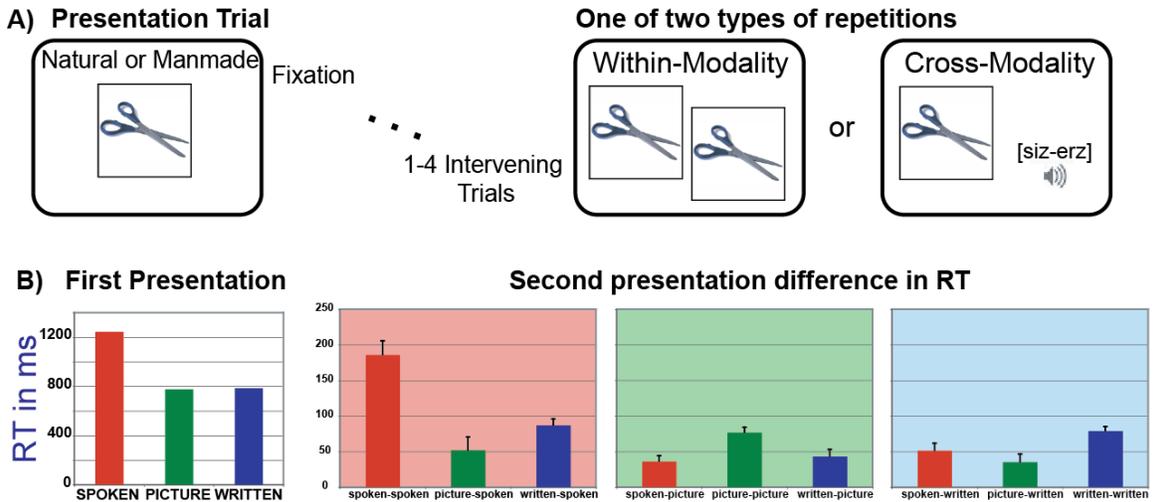

Figure 1:
A) Schematic of experimental design. On each trial participants were instructed to respond as quickly as possible with a natural or manmade judgment. Items were presented in one of three modalities (pictures, written words, spoken words). After 1-4 intervening trials, an item would repeat in either the original modality (perceptual), or one of the others (conceptual).
B) Response time in milliseconds for mean of first presentation in each modality. Difference in response time (in ms) for second presentations. Each condition showed significant behavioral priming ($p < .008$) Error bars denote standard error of the mean

**Figure 2**

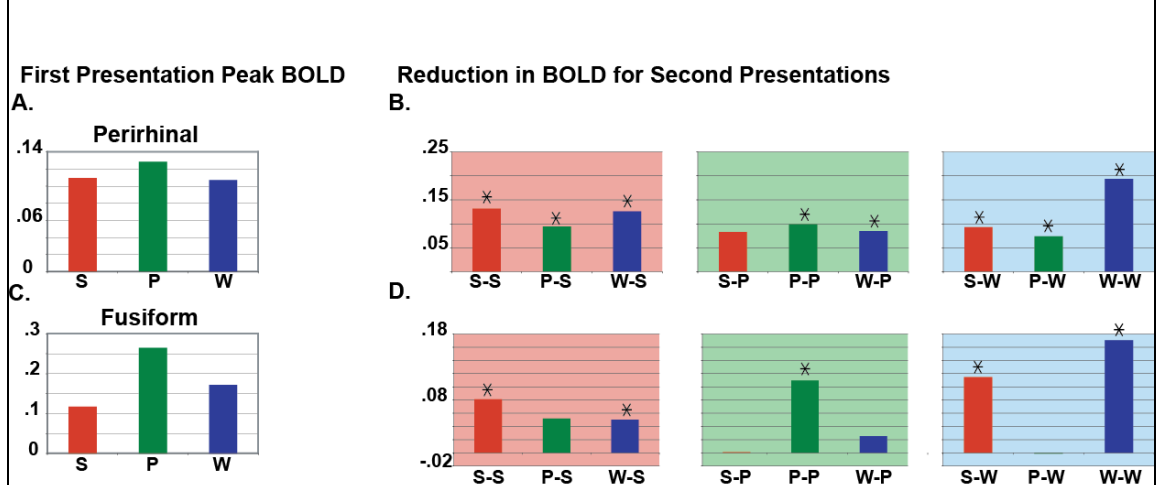

Figure 2: BOLD data
A,C: Peak BOLD signal (a.u.) correlated with first mean first presentation for a modality. S=spoken P= picture W= written
B,D: Reduction in peak BOLD signal (a.u.) for mean second presentation . * indicates significant reductions ($p<.05$, one tailed t-test)

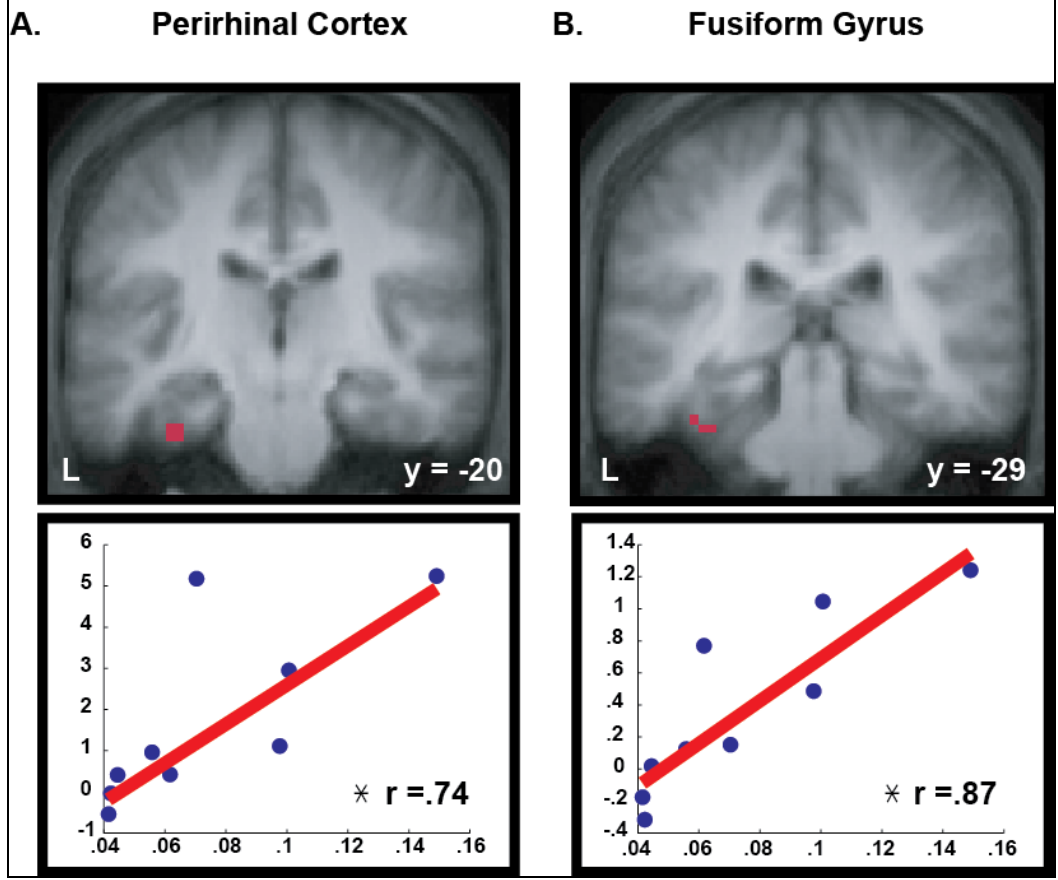

Figure 3: Correlation Data
Correlation between normalized behavioral priming (x axis) and normalized BOLD RS (y axis) across the nine repetition conditions; r value, Kendall's Tau. * denote significant ($p <.05$) correlations. Each condition denoted with a circle, line = regression line from robust fit regression. Activity in these regions is significantly correlated with conceptual processing.

**Supplemental Figure 1**

A

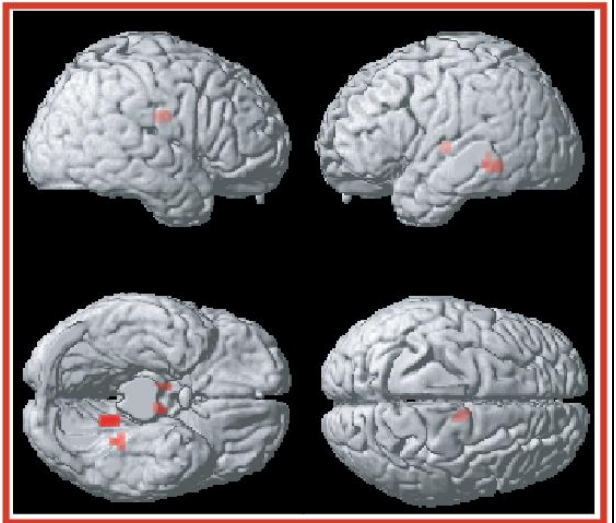

B

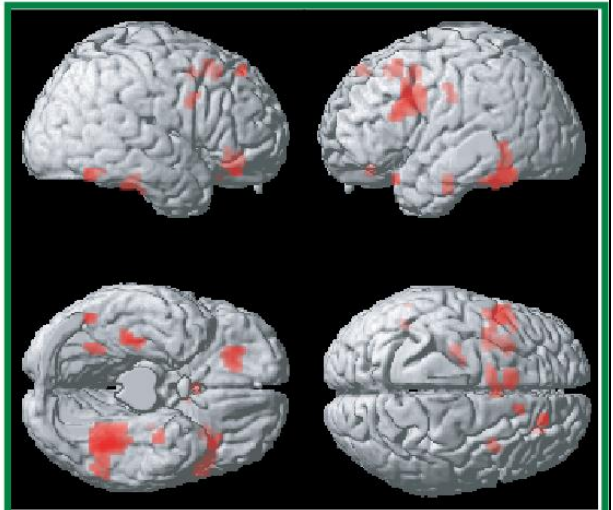

C

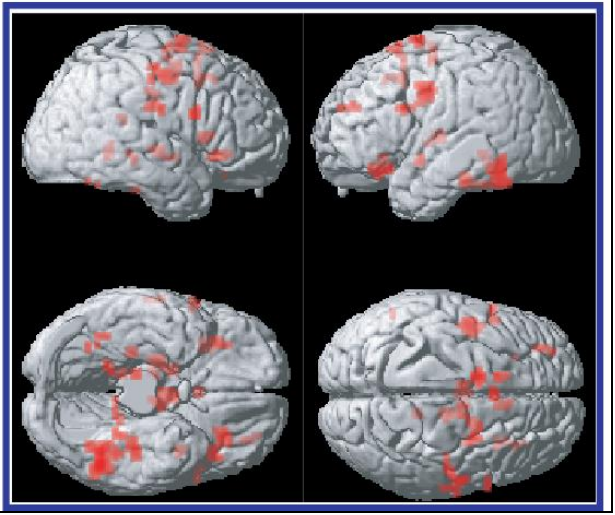

Supplemental Figure 1: Perceptual Repetition
Statistical parametric maps (*p* value of 0.005) of perceptual repetition rendered on a canonical brain
A) Spoken words B) Pictures C) Written words

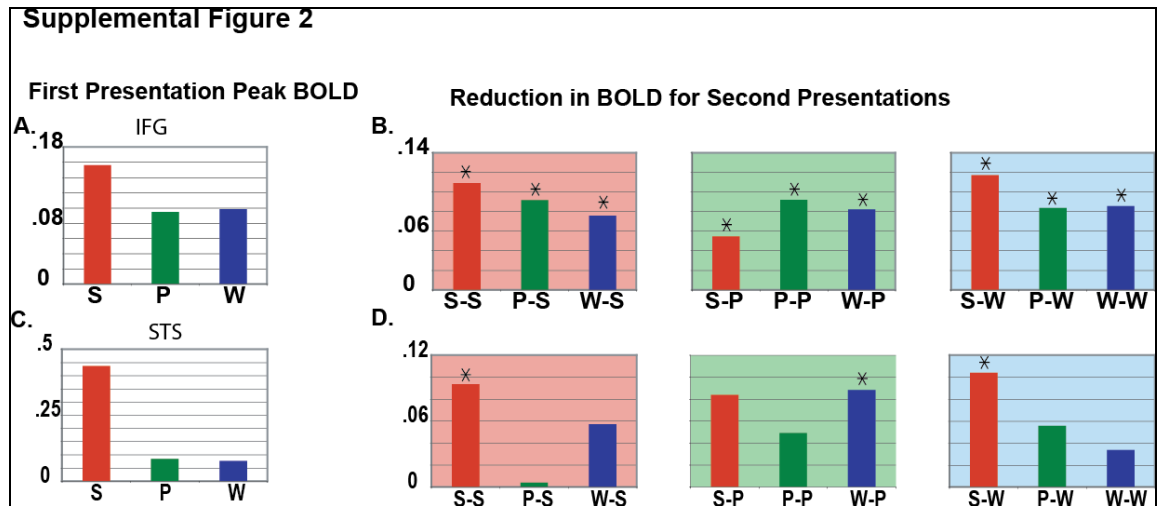

Supplemental Figure 2: BOLD data
A,C: Peak BOLD signal (a.u.) correlated with first mean first presentation for a modality. S=spoken P= picture W= written
B,D: Reduction in peak BOLD signal (a.u.) for mean second presentation . * indicates significant reductions ($p<.05$, one tailed t-test)

**Supplemental Figure 3**

A **Inferior Frontal Gyrus**

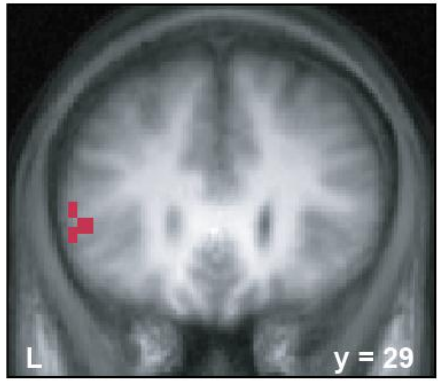

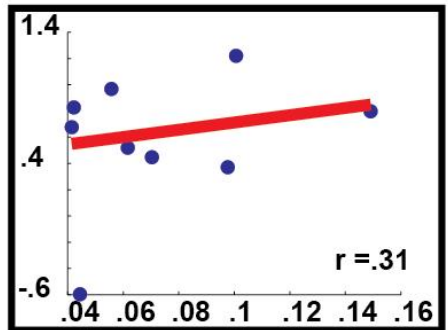

B **Superior Temporal Sulcus**

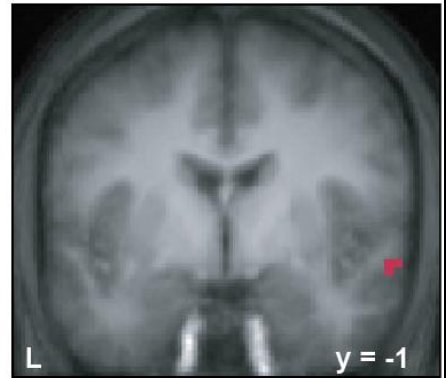

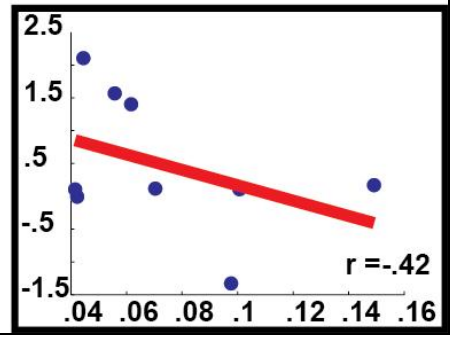

Figure 3: Correlation Data

Correlation between normalized behavioral priming (x axis) and normalized BOLD RS (y axis) across the nine repetition condition; r value, Kendall's Tau. * denote significant ($p <.05$) correlations. Each condition denoted with a circle, line = regression line from robust fit regression. Although both regions show repetition effects, in contrast with those in perirhinal and fusiform, these effects are not significantly correlated with behavior.

Citations

Bressler SL, Tang W, Sylvester CM, Shulman GL, Corbetta M (2008) Top-down control of human visual cortex by frontal and parietal cortex in anticipatory visual spatial attention. J Neurosci 28:10056–10061

Dale A M (1999) Optimal experimental design for event-related fMRI. Human Brain Mapping 8(2-3), 109-114

Deacon D, Hewitt S, Yang CM, Nagata M (2000) Event-related potential indices of semantic priming using masked and unmasked words: Evidence that the N400 does not reflect a post-lexical process. Cognitive Brain Research 9: 137–146

Dobbins IG, Schnyer DM, Verfaellie M, Schacter DL (2004) Cortical activity reductions during repetition priming can result from rapid response learning. Nature, 428: 316-319

Epstein R, Kanwisher N (1998) A Cortical Representation of the Local Visual Environment. Nature 392: 598-601

Ewbank MP, Schluppeck D, Andrews TJ (2005) fMR-adaptation reveals a distributed representation of inanimate objects and places in human visual cortex. NeuroImage 28:268-279

Fernández G, Tendolkar I (2006) The rhinal cortex: 'gatekeeper' of the declarative memory system. Trends in Cognitive Science 10: 358–362

Grelotti DJ, Klin AJ, Gauthier I, Skudlarski P, Cohen DJ, Gore JC, Volkmar FR, Schultz RT (2005) fMRI activation of the fusiform gyrus and amygdala to cartoon characters but not to faces in a boy with autism. Neuropsychologia 43(3): 373–385

Henson RN, (2003) Neuroimaging studies of priming. Progress in Neurobiology 70(1): 53–81

Henson RN, Goshen-Gottstein Y, Ganel T, Otten LJ, Quayle A, Rugg MD (2003) Electrophysiological and haemodynamic correlates of face perception, recognition and priming. Cerebral Cortex 13: 793–805

Holcomb PJ (1993) Semantic priming and stimulus degradation: Implications for the role of the N400 in language processing. Psychophysiology 30: 47-61

Holcomb PJ, Reder L, Misra M, Grainger J. (2005) Masked priming: an event-related brain potential study of repetition and semantic effects. Cognitive Brain Research, 24: 155-172.

Koutstaal W, Wagner AD, Rotte M, Maril A, Buckner RL, Schacter DL (2001) Perceptual specificity in visual object priming: functional magnetic resonance imaging evidence for a laterality difference in fusiform cortex. Neuropsychologia 39(2): 184-199

Lavenex P, Suzuki WA, Amaral DG (2002) Perirhinal and parahippocampal cortices of the macaque monkey: projections to the neocortex. Journal of Comparative Neurology 447: 394-420

O'Kane G, Insler RZ, Wagner A D (2005) Conceptual and perceptual novelty effects in human medial temporal cortex. Hippocampus 15: 326-332

Patterson K, Nestor PJ, Rogers T T (2007) Where do you know what you know? Nature Reviews Neuroscience 8: 976-988

Rissman J, Eliassen JC, Blumstein SE (2003) An event-related FMRI investigation of implicit semantic priming. Journal of Cognitive Neuroscience 15(8): 1160-1175

Rugg MD (1990) Event-related brain potentials dissociate repetition effects of high- and low-frequency words. Memory & Cognition 18: 367-379

Sayres R, Grill-Spector K (2006) Object-selective cortex exhibits performance-independent repetition suppression. Journal of Neurophysiology 95(2): 885-1007

Smith ME, Stapleton JM, Halgren E (1986) Human medial temporal lobe potentials evoked in memory and language tasks. Electroencephalogr Clin Neurophysiol 63:145-149

Supér H, Spekreijse H, Lamme V, (2001) Two distinct modes of sensory processing observed in monkey primary visual cortex (V1). Nature Neuroscience 4, 304-310

Suzuki WA, Amaral DG (1994) Perirhinal and parahippocampal cortices of the macaque monkey: cortical afferents. The Journal of Comparative Neurology 350: 497-533

Taylor KI, Moss HE, Stamatakis EA, Tyler LK (2006) Binding crossmodal object features in perirhinal cortex. Proc Natl Acad Sci U S A, 103(21): 8239-8244

Thiel CM, Henson RN, Morris JS, Friston KJ, Dolan RJ (2001) Pharmacological modulation of behavioral and neuronal correlates of repetition priming. Journal of Neuroscience 21(17): 6846-6852

Thiel A, Haupt WF, Habedank B, Winhuisen L, Herholz K, Kessler J, Markowitsch HJ, Heiss WD (2005) Neuroimaging guided rTMS of the left inferior frontal gyrus interferes with repetition priming. Neuroimage **25**:815–823

Vuilleumier P, Henson RN, Driver J, Dolan RJ (2002) Multiple levels of visual object constancy revealed by event-related fMRI of repetition priming. Nature Neuroscience, 5(5): 491-499

Wheatley T, Weisberg J, Beauchamp MS, Martin A (2005) Automatic priming of semantically related words reduces activity in the fusiform gyrus. Journal of Cognitive Neuroscience 17(12): 1871-1885

Wig GS, Grafton ST, Demos KE, Kelley WM (2005) Reductions in neural activity underlie behavioral components of repetition priming. Nature Neuroscience 8(9): 1228-1233